\documentclass[11pt]{iopart}

\usepackage{graphicx}
\usepackage{epstopdf}

\def\prl{Phys. Rev. Lett.}
\def\prd{Phys. Rev. D}
\def\cqg{Class. Quantum Grav.}

\def\apj{Astrophys. J.}

\def\pr{Phys. Rev.}

\begin{document}

\title{Oppenheimer-Snyder Collapse in Moving-Puncture Coordinates}

\author{A. N. Staley,$^1$ T. W. Baumgarte,$^{1,2}$ J. D. Brown,$^3$ B. Farris,$^2$ and S. L. Shapiro$^{2,4,5}$}

\address{${}^1$ Department of Physics and Astronomy, Bowdoin College,  
Brunswick, ME 04011, USA}

\address{${}^2$ Department of Physics, University of Illinois at
  Urbana-Champaign, Urbana, Illinois 61801, USA}

\address{${}^3$ Department of Physics, North Carolina State University, Raleigh, NC 27695, USA}

\address{${}^4$ Department of Astronomy, University
  of Illinois at Urbana-Champaign, Urbana, Illinois 61801, USA}
  
\address{${}^5$ NCSA, University
  of Illinois at Urbana-Champaign, Urbana, Illinois 61801, USA}

\begin{abstract}
Moving puncture coordinates are commonly used in numerical simulations of black holes.  Their properties for vacuum Schwarzschild black holes have been analyzed in a number of studies.  The behavior of moving-puncture coordinates  in spacetimes containing matter, however, is less well understood.  In this paper we explore the behavior of these coordinates for Oppenheimer-Snyder collapse, i.e., the collapse of a uniform density, pressureless sphere of dust initially at rest to a black hole.  Oppenheimer-Snyder collapse provides a stringent test of the singularity-avoiding properties of moving-puncture coordinates, since the singularity can form more quickly than it would for matter with pressure.  Our results include analytical expressions for the  matter density, lapse function, and mean curvature at early times, as well as interesting limits for later times.  We also carry out numerical simulations to obtain the full solution and these show that, even in the absence of pressure, moving-puncture coordinates are able to avoid the singularity. At late times the geometry settles down to a trumpet slice of a vacuum black hole. 
\end{abstract}

\section{Introduction}

Numerical simulations of black holes have recently experienced a dramatic breakthrough (see \cite{Pre05b,BakCCKM06a,CamLMZ06} as well as numerous later publications).   Many of these simulations now adopt some variation of the BSSN formulation \cite{ShiN95,BauS99} together with moving-puncture coordinates \cite{BakCCKM06a,CamLMZ06} to handle the black hole singularities (see also \cite{BauS10} for a review).

Moving-puncture coordinates adopt the 1+log slicing condition for the lapse function \cite{BonMSS95} and a gamma-driver condition for the shift vector \cite{AlcBDKPST03}.  The properties of moving-puncture coordinates have been studied in detail for Schwarzschild spacetimes \cite{HanHPBO06,HanHOBGS06,Bro08,HanHOBO08,DenWBB10}.  In particular, these studies show that dynamical evolutions of Schwarzschild black holes with moving-puncture coordinates settle down to a slice that ends on a sphere of finite areal radius.  When displayed in an embedding diagram (see, e.g., Figure 2 in \cite{HanHOBO08}), these slices resemble a trumpet, which motivates the name ``trumpet geometry".  The properties of these slices, in particular the fact that they do not extend to the singularity but instead end at finite areal radius, explain why these slices are well suited for dynamical simulations.

Studying interactions of matter and black holes, which is important for many problems of astrophysical interest, requires relaxing the assumption of pure vacuum and including matter sources in the numerical simulations.   While numerous simulations have already adopted moving-puncture coordinates in evolution calculations involving matter (see, e.g., \cite{FabBEST07,BaiR06,MonFS08,etienne09,ThiBB11}), the properties of these coordinates in non-vacuum spacetimes are less well understood than in vacuum spacetimes.

In a recent study, Thierfelder \etal \cite{ThiBHBR11} analyzed the collapse of a spherically symmetric, unstable fluid star using two variations of moving-puncture coordinates (see their equation (2)).   In the version that is similar to the gamma-driver condition most commonly used in binary black hole simulations ($\mu_S = 1$ in their equation (2); see also \cite{MetBKC06}), the grid-points are pulled out of the star at late times during the collapse (see also \cite{Shi99a} as well as the discussion at the end of Chapter 4 of \cite{BauS10}).  All matter is effectively removed from the computational grid, and the simulation settles down to a trumpet geometry that becomes indistinguishable from that of a vacuum black hole.  During this process the stellar interior becomes poorly resolved; this can make the code crash unless certain hydrodynamical values are prevented from taking unphysical values (see \cite{ThiBHBR11} for details).   In an alternative version of the gamma-driver shift condition ($\mu_S = \alpha^2$ in their equation (2)), the grid points are not pulled out of the star. Instead, the grid-points are ``frozen" at late times and remain in the stellar interior; this approach leaves the matter well resolved even at late times.

A classical, analytical solution describing the collapse of matter into a black hole is Oppenheimer-Snyder (OS) collapse of a homogeneous dust sphere \cite{OppS39}.  This solution has served as a testbed for numerous numerical codes, and it has proven useful to transform this solution, originally derived in Gaussian normal coordinates, into coordinate systems that are more commonly used in numerical simulations (including maximal slicing \cite{PetST85}, polar slicing \cite{PetST86} and ``observer-time" coordinates \cite{BauST95}).  It is therefore of interest to study OS collapse in moving-puncture coordinates as well, which we do below.

Moving-puncture coordinates form a rather complicated set of equations. It appears impossible to obtain a complete analytical solution to the OS collapse problem in these coordinates.  However, we can obtain some partial analytical results and we do so here to illuminate our numerical integrations.  Specifically, we find that at early times the matter density, lapse function, and mean curvature remain constant in space while varying in time.  We also show that these functions remain spatially constant until a gauge mode has propagated from the stellar surface.  Finally, we derive a lower limit for the lapse function at the center of the dust cloud. 

OS collapse poses another interesting question: can the lapse in moving-puncture coordinates collapse sufficiently fast for the slicing to avoid the black hole singularity, and to exhibit a transition to a trumpet geometry?  As demonstrated by \cite{ThiBHBR11}, this is the case for collapse of a fluid star with pressure.  In the absence of pressure, however, as in OS collapse, the matter collapses more rapidly, leaving the lapse less time to drop to zero and avoid the singularity.  In this sense, OS collapse poses a more stringent test than the collapse of a fluid star with pressure.  We demonstrate numerically that, even for OS collapse, moving-puncture coordinates do avoid the singularity and  allow for a transition to a trumpet geometry.

This paper is organized as follows.  In Section \ref{OS} we review the OS solution and introduce a transformation to general coordinates.  In Section \ref{early_times} we specialize to the 1+log slicing condition. We show that the density, lapse, and mean curvature remain spatially constant at each point of the stellar interior until that point is reached by a gauge mode propagating from the stellar surface.  In Section \ref{late_times} we present the full numerical solution. We use this solution to probe the late-time behavior of OS collapse, demonstrating the transition to a trumpet geometry.  We briefly summarize our results in Section \ref{sum}.  Throughout this paper we use geometrized units in which $G = c = 1$.

\section{Oppenheimer-Snyder collapse}
\label{OS}

\subsection{Solution in Gaussian normal coordinates}

\begin{figure}
\includegraphics[width=3in]{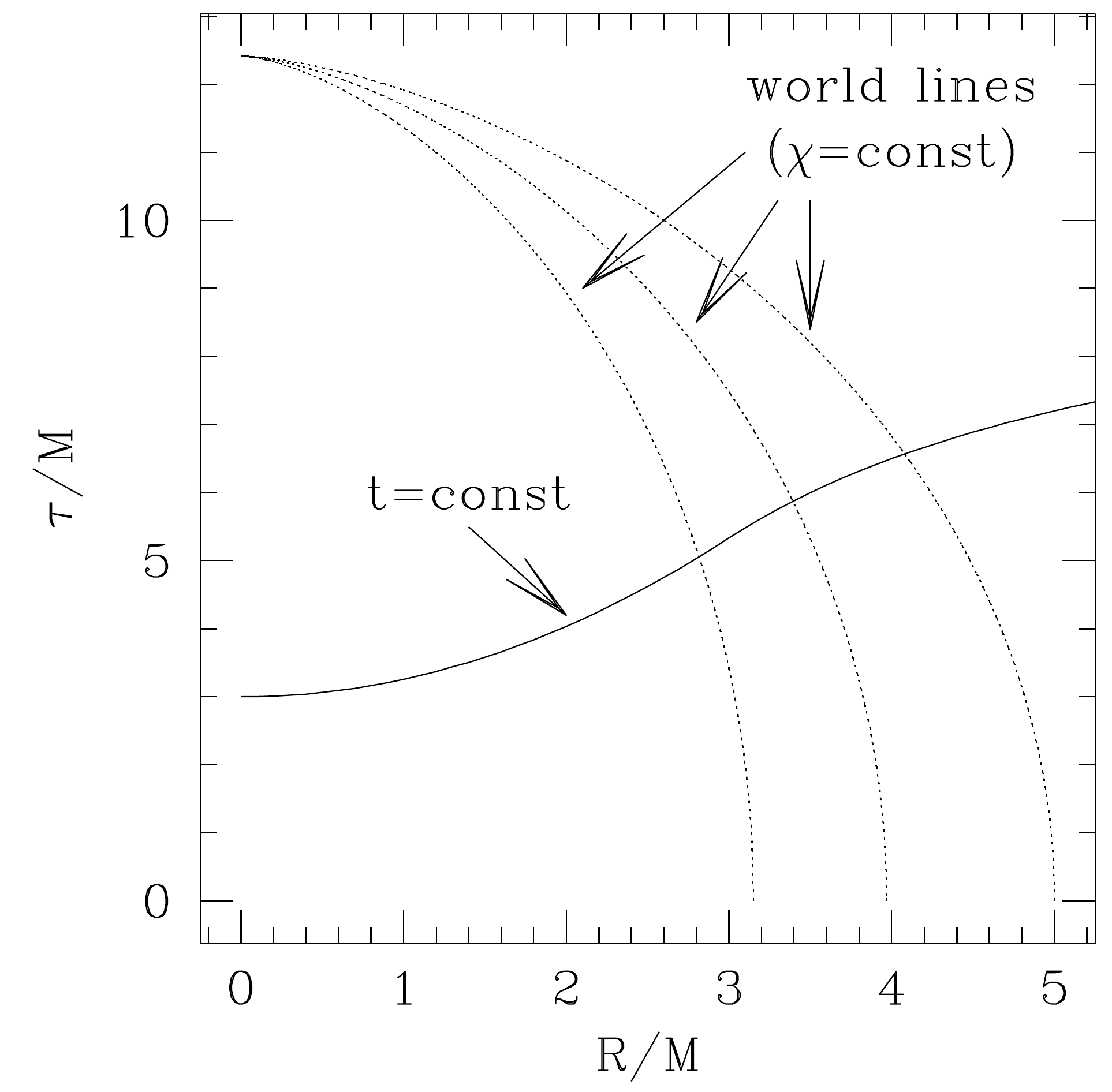}
\caption{Spacetime diagrams for OS collapse, for an initial areal radius of $R_0 = 5 M$.  The dotted lines show trajectories of dust particles, which correspond to constant values of the radial coordinate $\chi$.  The solid line represents a hypothetical surface of constant coordinate time $t$.  }
\label{fig1}
\end{figure}

OS collapse \cite{OppS39} describes the collapse from rest of a constant-density dust sphere to a black hole  (see Figure \ref{fig1} for a spacetime diagram, see also Section 1.4 in \cite{BauS10} for a review).  The solution can be given analytically by matching a closed-Friedmann solution for 
the stellar interior to a Schwarzschild solution for the exterior. The interior solution, expressed in Gaussian normal coordinates, is given by
\begin{equation} \label{met_gn}
ds^2 = - d\tau^2 + a^2(\tau) \left(d\chi^2 + \sin^2 \chi d\Omega^2\right),
\end{equation}
where $0 \leq \chi \leq \chi_0$ is a radial coordinate that is comoving with each dust particle.  
The scale factor $a(\tau)$ can be expressed parametrically as a function of the proper time as measured by each dust particle with the help of the conformal time $\eta$,
 \begin{equation} \label{friedmann_parameters}
 \begin{array}{rcl}
 a & = & \displaystyle \frac{1}{2} \, a_{\rm m} (1 + \cos \eta), \\[3mm]
 \tau & = & \displaystyle \frac{1}{2} \, a_{\rm m} (\eta + \sin \eta). 
\end{array}
\end{equation}
Matching this solution to an exterior Schwarzschild solution shows that the initial scale factor $a_{\rm m}$ and the maximum value $\chi_0$ of the radial coordinate $\chi$ are given by
\begin{equation}  \label{a_m}
a_{\rm m} = \left( \frac{R_0^3}{2 M} \right)^{1/2}
\end{equation} 
and
\begin{equation} \label{chi_0}
\chi_0 = \sin^{-1}\left( \sqrt{\frac{2M}{R_0}} \right) ,
\end{equation}
where $R_0$ is the initial areal radius of the cloud, and $M$ is the cloud's ADM mass.  

The rest-mass density $\rho_0$, as observed by an observer comoving with the matter, remains homogeneous (i.e.~spatially constant) on each slice of constant $\tau$, and is given by
\begin{equation} \label{density}
\frac{\rho_0(\tau)}{\rho_0(0)} = \left( \frac{a_{\rm m}}{a(\tau)} \right)^3.
\end{equation}
Here the initial density at $\tau = 0$,  $\rho_0(0)$, is related to the initial areal radius $R_0$ and the mass $M$ by
\begin{equation}
M = \frac{4\pi}{3} \rho_0(0) R_0^3.
\end{equation}
We note that, for dust, $\rho_0$ is equal to the total mass-energy density as observed by a comoving observer, so that $\rho_0 = u_a u_b T^{ab}$, where $u^a$ is the dust particles's four-velocity and $T^{ab}$ the stress-energy tensor.

All particles reach the singularity with infinite density $\rho_0(\tau)$ when $a = 0$.   According to equation (\ref{friedmann_parameters}) this corresponds to a conformal time of $\eta = \pi$, or a proper time
\begin{equation} \label{tau_sing}
\frac{\tau_{\rm sing}}{M} = \left( \frac{R_0}{2 M} \right)^{3/2} \pi.
\end{equation}

In Gaussian normal coordinates, the mean curvature $K = \gamma^{ij} K_{ij}$, where $\gamma_{ij}$ is the spatial metric and $K_{ij}$ the extrinsic curvature, is given by
\begin{equation} \label{K_GN}
K_{\rm GN} = - \frac{3}{a} \, \frac{d a}{d\tau}.
\end{equation}
We also note that the ``Gaussian normal" normal vector $n^a_{\rm GN}$, which is orthogonal on slices of constant proper time $\tau$, is aligned with the dust particles' four-velocity $u^a$,
\begin{equation}
n_{\rm GN}^a = u^a.
\end{equation}

\subsection{General coordinate transformation}

We first consider a general coordinate transformation to a new time coordinate $t$ and a new radial coordinate $r$.   We have included a hypothetical surface of constant $t$ in Figure~\ref{fig1}.  We assume that we can express the old coordinates $\tau$ and $\chi$ as functions of the new coordinates,
\begin{equation}
\tau = \tau(t,r) ~~~~\mbox{and}~~~~ \chi=\chi(t,r);
\end{equation}
we will also assume that $t=0$ for $\tau = 0$ and $r = 0$ at the origin $\chi = 0$.  Inserting these functions into the line element (\ref{met_gn}) results in the new line element
\begin{equation} \label{met_trans}
\begin{array}{rcl}
ds^2 & = & - (\dot \tau^2 - a^2 \dot \chi^2) dt^2 + 2 (a^2 \dot \chi \chi' - \dot \tau \tau') dr dt \\
& & + (a^2 \chi'^2 - \tau'^2) dr^2 + a^2 \sin^2 \chi d \Omega^2,
\end{array}
\end{equation}
where we have defined the partial derivatives
\begin{equation}
\begin{array}{rclrcl}
 \dot \tau & \equiv & \displaystyle \left. \frac{\partial \tau}{\partial t} \right|_r ~~~~~~~&
 \tau' & \equiv & \displaystyle \left. \frac{\partial \tau}{\partial r} \right|_t \\[5mm]
  \dot \chi & \equiv & \displaystyle \left. \frac{\partial \chi}{\partial t} \right|_r  &
 \chi' & \equiv & \displaystyle \left. \frac{\partial \chi}{\partial r} \right|_t .
\end{array}
\end{equation}
Comparing the line element  (\ref{met_trans}) with the $3+1$ form of the line element
\begin{equation} \label{met_3+1}
ds^2 = - \alpha^2 dt^2 + \gamma_{ij} (dx^i + \beta^i dt)(dx^j + \beta^j dt)
\end{equation}
we first identify the spatial metric
\begin{equation} \label{spat_met}
\gamma_{ij} = \mbox{diag}(a^2 \chi'^2 - \tau'^2, a^2 \sin^2 \chi, a^2 \sin^2 \chi \sin^2 \theta)
\end{equation}
and its determinant
\begin{equation} \label{det}
\gamma = (a^2 \chi'^2 - \tau'^2) a^4 \sin^4 \chi \sin^2 \theta.
\end{equation}
We can then read off the radial, and only non-vanishing, component of the shift vector
\begin{equation} \label{shift}
\beta_r = a^2 \dot \chi \chi' - \dot \tau \tau'  ~~~~~\mbox{or}~~~~
\beta^r = \frac{a^2 \dot \chi \chi' - \dot \tau \tau'}{a^2 \chi'^2 - \tau'^2}.
\end{equation}
Finally, the lapse function is given by
\begin{equation} \label{lapse}
\alpha^2 = \frac{a^2}{\tau'^2 - a^2 \chi'^2} \left( \dot \tau \chi' - \dot \chi \tau' \right)^2.
\end{equation}

\section{Early stage of the collapse: analytical results}
\label{early_times}

\subsection{The 1+log condition}

We now impose the 1+log slicing condition
\begin{equation} \label{1+log}
(\partial_t - \beta^i \partial_i) \alpha = - 2 \alpha K.
\end{equation}
Here the mean curvature $K = \gamma^{ij} K_{ij}$ can be expressed as
\begin{equation} \label{K}
K = - \frac{1}{2\alpha} \left(\frac{\dot \gamma}{\gamma} - 2 \frac{1}{\sqrt{\gamma}}
	\partial_r \left(\sqrt{\gamma} \beta^r\right) \right).
\end{equation}
We will also use the time evolution equation for $K$,
\begin{equation} \label{K_dot}
(\partial_t - \beta^i \partial_i) K = - \gamma^{ij} D_i D_j \alpha + \alpha K_{ij}K^{ij} + 4 \pi \alpha (\rho + S),
\end{equation}
where $D_i$ denotes the covariant derivative associated with the spatial metric $\gamma_{ij}$,
$K_{ij}$ is the extrinsic curvature, $\rho \equiv n_a n_b T^{ab}$ is the matter density as observed by a normal observer, and $S \equiv \gamma^{ab} \gamma_a{}^c T_{bc}$ is the trace of the spatial stress, again as observed by a normal observer.

We note that the slicing condition (\ref{1+log}) can also be written as
\begin{equation} \label{1+log_normal}
\frac{d}{d\tau_n} \alpha = n^a \partial_a \alpha = - 2 K,
\end{equation}
where $n^a$ is the normal vector on each slice of constant coordinate time $t$ (which, in general, is different from $n_{\rm GN}^a$), and where $\tau_n$ is the proper time as observed by such a normal observer.  This shows that 1+log slicing is invariant under spatial coordinate transformations, i.e.~independent of the choice of the shift (see also \cite{Alc03}).   In order to find the lapse $\alpha$, we may therefore choose whatever shift is convenient, and we will consider two different choices below.

We also observe that the pair of equations formed by the 1+log slicing condition (\ref{1+log}) and the evolution equation for the mean curvature (\ref{K_dot}) can be combined to form a wave equation for the lapse.   For simplicity we may assume vanishing shift.  Taking a time derivative of equation (\ref{1+log}) and using the result to eliminate the term $\partial_t K$ in (\ref{K_dot}) we obtain an inhomogeneous wave equation of the form
\begin{equation} \label{lapse_wave}
-  \partial^2_t \alpha + 2 \alpha \gamma^{ij} D_i D_j \alpha = \ldots ,
\end{equation}
where the right hand side contains lower-order terms that do not affect the wave operator on the left-hand side. 

\subsection{Spatially constant lapse}

We first show that if $\alpha$, $K$ and $\rho_0$ are spatially constant in a region of a spatial slice, then they will remain spatially constant in the domain of dependence of this region.  These quantities may then change in time, but not in space.  This result can be verified directly from the equations; it is easier, however, to see this from the properties of OS collapse.  

According to equation (\ref{density}), slices of constant $\rho_0$ in OS collapse correspond to slices of constant proper time $\tau$ (which is measured along the trajectories of dust particles).  The region that we are considering therefore corresponds to a region of a slice of constant $\tau$.  Moreover, the normal on these slices is aligned with the four-velocity of the dust particles, $n^a = u^a$, which implies $\rho = \rho_0$.  According to equation (\ref{K_GN}), $K$ is also constant on such a slice. We now assume a more general, but still spatially constant, lapse.  Integrating forward to a new time slice of constant coordinate time $t + dt$, the proper time as measured along the slice's normal (and hence along the dust particle's trajectory) will increase by $d \tau = \alpha dt$.   In a region where $\alpha$ is spatially constant, the new slice will again correspond to a slice of constant $\tau$.  Moreover, with $K$ constant in this region, equation (\ref{1+log_normal}) shows that the lapse will again be constant on the new slice.

In particular, this argument applies to early times, assuming the integration starts with a constant initial lapse $\alpha_0$.   On the other hand, this argument does not apply at the surface of the star.  At the star's surface, the density $\rho$ is discontinuous, and since this quantity appears on the right hand side of (\ref{K_dot}), the mean curvature will not remain spatially constant.  We therefore expect that the surface will trigger a deviation that propagates into the star, satisfying the wave equation (\ref{lapse_wave}).  The quantities $\rho$, $K$ and $\alpha$ will remain spatially constant at any location in the star only until this deviation has arrived; i.e.~in the domain of dependence of the initial stellar interior.   We show a numerical demonstration of this behavior in Figure \ref{fig2}.

Assuming a spatially constant lapse we can obtain its analytic value by choosing $\beta^r = 0$.  Equation (\ref{1+log}) then simplifies and can be integrated immediately to yield
\begin{equation}
\alpha = \alpha_0 + \ln(\gamma/\gamma_0),
\end{equation}
where the subscript 0 denotes initial values at $\tau = 0$.   Inserting the determinant (\ref{det}) we find
\begin{equation} \label{lapse_general}
\alpha = \alpha_0 + \ln\left( \frac{(a^2 \chi'^2 - \tau'^2) a^4}{(a_{\rm m}^2 \chi_0'^2 - \tau_0'^2) a_{\rm m}^4}\right).
\end{equation}
It is difficult to make further progress in general.  If we assume, however, that the lapse has remained spatially constant, as it does at early times, then the above expression simplifies.   While the lapse remains constant, slices of constant time $t$ will coincide with slices of constant $\tau$, meaning that $\tau' = 0$ and that $n^a = n_{\rm GN}^a$.  For zero shift, coordinate observers are comoving with normal observers; this means that the new coordinates $r$ are comoving with the fluid, which, in turn, are described by fixed values of the original radial coordinate $\chi$.  For zero shift, and as long as the lapse remains constant, we therefore have $\chi' = \chi_0'$, and the lapse (\ref{lapse_general}) reduces to
\begin{equation} \label{lapse_const}
\alpha = \alpha_0 + 6 \ln (a/a_{\rm m})~~~~\mbox{(as long as $\alpha$ remains spatially constant)}.
\end{equation}
Alternatively, this expression can be derived by observing that, on slices of constant $\tau$, the mean curvature is given by (\ref{K_GN}), so that for vanishing shift equation (\ref{1+log}) can be integrated directly to yield (\ref{lapse_const}).

\subsection{The lapse at the center}

We will now show that the expression on the right--hand side of Eq.~(\ref{lapse_const}) is a lower limit for the lapse function. 
To begin, we choose the shift vector in such a way that $\chi = r$ at all times.  In this case we have $\dot \chi = 0$ and $\chi' = 1$, so that the lapse reduces to
\begin{equation} \label{lapse2}
\alpha^2 = \frac{a^2 \dot \tau^2}{a^2 - \tau'^2}
\end{equation}
and the shift to
\begin{equation}
\beta^r = - \frac{\dot \tau \tau'}{a^2 - \tau'^2}.
\end{equation}
Regularity at the origin demands $\beta^r = 0$ there, which implies $\tau' = 0$ at $r = 0$ (as long as $\dot \tau \neq 0$).  This means that slices of constant coordinate time $t$ must be tangent to slices of constant $\tau$ at the origin, as indicated in Figure \ref{fig1}.  Inserting the lapse and shift, together with (\ref{K}), into the 1+log slicing condition (\ref{1+log}), and evaluating the resulting expression at the origin $r=0$, we obtain
\begin{equation} \label{taudotdot}
\ddot \tau - 6 \frac{1}{a} \frac{da}{dt} - 2 \frac{\tau'' \dot \tau}{a^2} = 0.
\end{equation}
This is a complicated equation for $\tau$, and it is again difficult to make progress, even after having restricted the analysis to the origin.  However, we can derive a useful limit from the above expression.  

\begin{figure}
\includegraphics[width=3in]{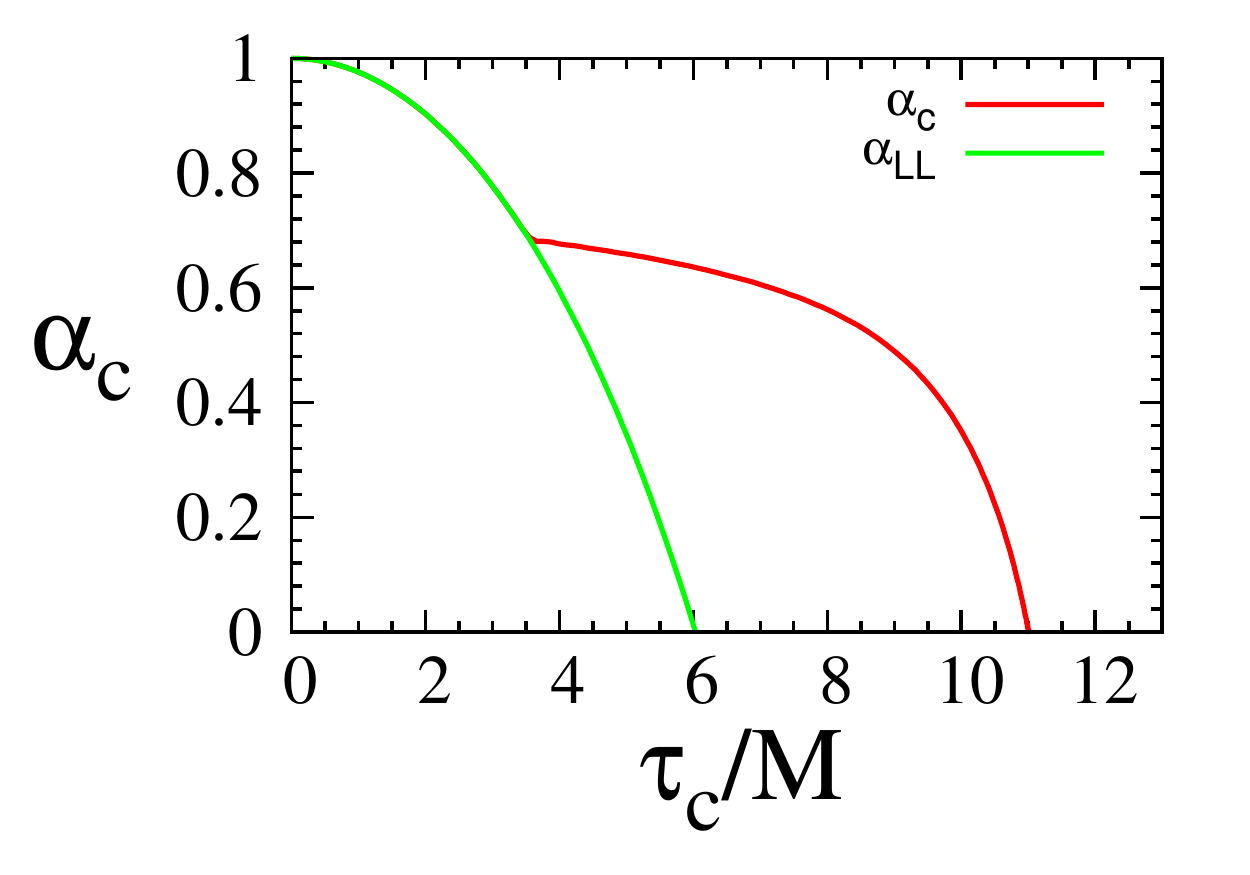}
\includegraphics[width=3in]{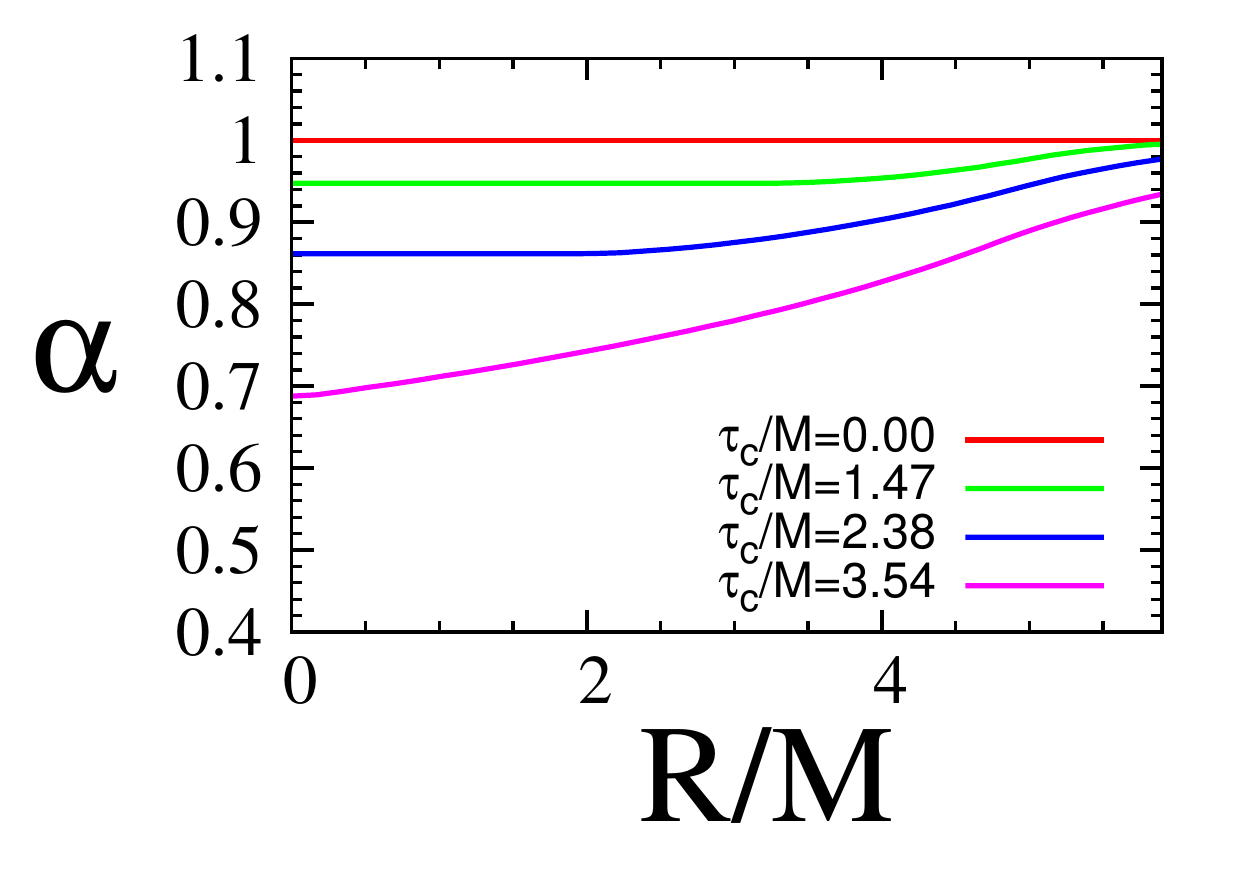}
\caption{Values for the lapse $\alpha$ for OS collapse with an initial areal radius of $R_0 = 5 M$.  The left panel shows the values of the lapse at the center, $\alpha_{\rm c}$, together with the lower limit $\alpha_{\rm LL}$ (see equation (\ref{lapse_ll})).  Note that the lapse (red) agrees with the analytical lower limit (green) at early times, but suddenly starts deviating from $\alpha_{\rm LL}$ at a time $\tau_{\rm gauge} \approx 3.5 M$.   The right panel shows profiles of the lapse as a function of areal radius $R$ at different times $t$, parametrized by the proper time $\tau_{\rm c}$ at the center. Red, green, blue and magenta lines correspond to $t/M=$ 0, 1.47, 2.38 and 3.54 respectively. At early times, the lapse remains spatially constant in a region decreasing in size around the center.  At the time $\tau_{\rm gauge}$ this region disappears, and the lapse is no longer constant.}
\label{fig2}
\end{figure}

Let us assume that the lapse has never taken a local maximum at the center; in other words, we assume that the lapse has always been either spatially constant in a neighborhood of the center, or it has taken a local minimum.  This condition holds at early times, while the lapse remains spatially constant, but even at later times this is a reasonable assumption for a slicing condition with ``singularity-avoiding" properties; this is also what we find in our numerical simulations of OS collapse in moving-puncture coordinates.   Under these conditions, coordinate slices of constant $t$ will be ``held back" at the center, as suggested in Figure \ref{fig1},  and we will have
\begin{equation}
\tau'' \geq 0
\end{equation}
at the center.  We can then integrate equation (\ref{taudotdot}) to obtain the limit for $\dot \tau \ge \dot\tau(0) + 6\ln(a/a_m)$. Moreover, equation (\ref{lapse2}) demonstrates that $\alpha = \dot \tau$ at the center; hence
\begin{equation} \label{lapse_ll}
\alpha_{\rm c} \geq \alpha_0 + 6 \ln(a/a_{\rm m}) \equiv \alpha_{\rm LL}.~~~~\mbox{(as long as $\tau''$ remains positive)} 
\end{equation}
Not surprisingly, the lower limit $\alpha_{\rm LL}$ agrees with the value that we obtained under the assumption of spatially constant lapse in equation (\ref{lapse_const}).  

In Figure \ref{fig2} we show numerical results that confirm the above findings.   These simulations were performed with a code that we describe in more detail in Section \ref{late_times} below.  In the left panel of Figure \ref{fig2} we show both $\alpha_{\rm c}$ and $\alpha_{\rm LL}$ as a function of proper time.   At early times, while the lapse remains spatially constant so that $\tau'' = 0$, the two agree as expected.   In the right panel of Figure \ref{fig2} we show profiles of the lapse at different times $t$.  These profiles demonstrate that the lapse indeed  remains spatially constant in a region around the center.   At a time that, for reasons that will become clear below, we will call the ``gauge time" $\tau_{\rm gauge}$, the central lapse suddenly starts to deviate from its lower limit.  For $R_0 = 5 M$ as in Figure \ref{fig2} we have $\tau_{\rm gauge} \approx 3.5 M$.  From the right panel in Figure \ref{fig2} we see that the region of spatially constant lapse disappears at the same time.  
\subsection{The gauge time $\tau_{\rm gauge}$}
\label{tau}

We next analyze the gauge time $\tau_{\rm gauge}$.  The numerical results in Figure \ref{fig2} suggest that this time might correspond to the time at which the center has come into causal contact with the surface. 

\begin{figure}
\includegraphics[width=3in]{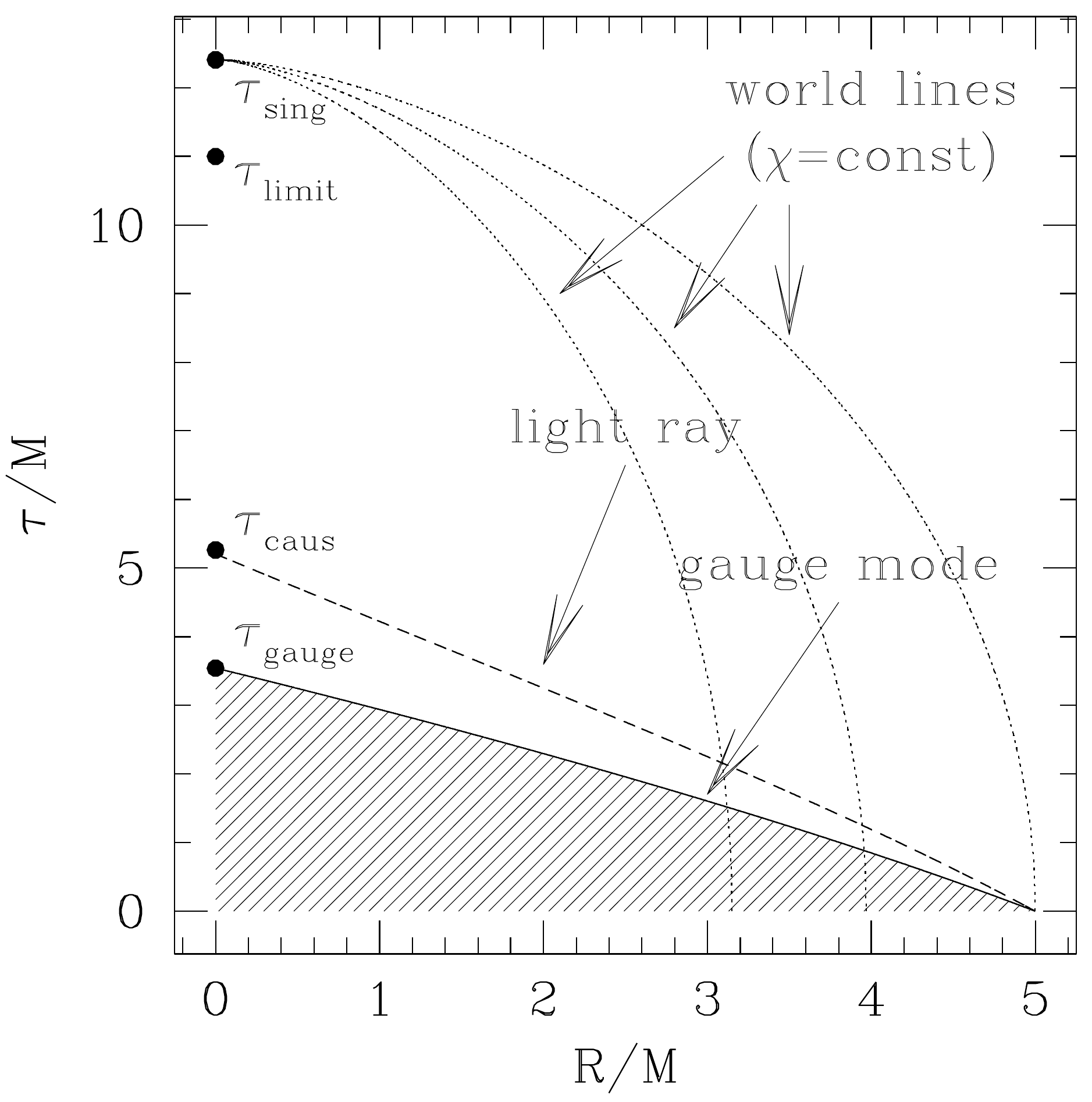}
\caption{A spacetime diagram of OS collapse for $R_0 = 5 M$ as in Figure \ref{fig1}, except that here we include two characteristic curves originating from the stellar surface at $\tau = 0$.  The dashed line marks a null characteristic, i.e.~the trajectory of a light ray, while the solid line marks a superluminal gauge mode propagating into the star.  As explained in the text, the density $\rho$, the lapse $\alpha$ and the mean curvature $K$ remain spatially constant in the shaded region below the solid line, which has not yet come into ``gauge" contact with the surface.  Also included are four critical times at the center of the star, namely the time $\tau_{\rm gauge}$ at which the center comes into ``gauge" contact with the surface, $\tau_{\rm caus}$ when it comes into causal contact with the surface, $\tau_{\rm limit}$ when the 1+log lapse drops to zero and halts any further evolution, and the proper time $\tau_{\rm sing}$ at which the singularity forms.}
\label{fig4}
\end{figure}

We first compute the time at which a light ray, emitted from the surface of the star at $t = 0$, reaches the center.   From equations (\ref{friedmann_parameters}) we have $d \tau = a \, d \eta$, so that we can rewrite the Friedmann metric (\ref{met_gn}) in terms of the conformal time $\eta$ in the conformal form
\begin{equation}
ds^2 =  a^2(\tau) \left( - d\eta^2 + d\chi^2 + \sin^2 \chi d\Omega^2 \right)
\end{equation}
(this motivates the name ``conformal time" for $\eta$).  Evidently, an ingoing, radial light ray satisfies $d\eta = - d\chi$.  A light ray sent out at conformal time $\eta = 0$ from the surface of the star at  $\chi = \chi_0$ reaches the center at a conformal time $\eta = \chi_0$, or at a proper time
\begin{equation}
\tau_{\rm caus} = \frac{1}{2} a_{\rm m} ( \chi_0 + \sin \chi_0 ).
\end{equation}
Inserting equations (\ref{a_m}) and (\ref{chi_0}) we can write this in terms of $R_0$ as
\begin{equation} \label{tau_caus}
\frac{\tau_{\rm caus}}{M} = \left(\frac{R_0}{2M} \right)^{3/2} \left( \sin^{-1} \left(\frac{2M}{R_0}\right)^{1/2} + 
\left(\frac{2M}{R_0}\right)^{1/2} \right),
\end{equation}
which, in the weak-field limit $2M/R_0 \rightarrow 0$, yields the expected result $\tau_{\rm caus} \rightarrow R_0$.  We have included this null characteristic as a dashed line in Figure \ref{fig4}.  

Evaluating (\ref{tau_caus}) for $R_0 = 5 M$ we find $\tau_{\rm caus} = 5.21 M$ (see Figure \ref{fig4}), which is {\em later} than $\tau_{\rm gauge}$.   This is not necessarily surprising, since the speed of light will only affect physical quantities.  The lapse, however, is a gauge quantity, which may propagate at speeds faster than the speed of light.  An analysis of the characteristic speeds for the BSSN system in moving-puncture coordinates shows that some of them are indeed faster than the speed of light \cite{Beyer:2004sv,Bro09}.  For sufficiently weak gravitational fields the fastest characteristic, labelled $\chi_4$ in \cite{Bro09}, has a proper speed, as measured by a normal observer whose four-velocity coincides with the normal on the slicing $n^a$, of
\begin{equation} \label{v_prop}
v_{\rm char} = - \sqrt{2/\alpha}
\end{equation}
(see Table I in \cite{Bro09}).

At least heuristically, the existence of this characteristic speed can be motivated quite easily from the wave equation 
(\ref{lapse_wave}).   From the principal part on the left hand side of this equation we find that radial modes propagate at coordinate speeds
\begin{equation}
v_{\rm char}^{\rm coord} = \pm(2 \alpha \gamma^{rr})^{1/2}.
\end{equation}
Following the procedure described in \cite{Bro09} these coordinate speeds can be converted into the proper speed (\ref{v_prop}).  

A normal observer comoving with the matter in the interior solution of OS collapse measures proper distances $a d\chi$ and proper times $d\tau = a d\eta$; such an observer therefore observes this gauge mode propagating at a speed
\begin{equation} \label{gauge_char1}
\frac{a d\chi}{d \tau} = \frac{d \chi}{d\eta} = -\sqrt{\frac{2}{\alpha}}.
\end{equation}
We now assume that the lapse remains spatially constant in a region that has not yet come into ``gauge" contact with the stellar surface; in this case the lapse is given by equation (\ref{lapse_const}) and we can rewrite (\ref{gauge_char1}) as 
\begin{equation}
d\chi = -\left[\frac{1}{2}  + 3 \ln \left(\frac{1 + \cos \eta}{2}\right) \right]^{-1/2} d \eta,
\end{equation}
where we have assumed $\alpha_0 = 1$.  It does not seem possible to evaluate this integral analytically.  A simple numerical integration from $\chi_0$ to $\chi = 0$ shows that, for $R_0 = 5 M$, this gauge mode arrives at the center at a conformal time $\eta_{\rm gauge} = 0.456$, or a proper time 
$\tau_{\rm gauge} = 3.54 M$.  This agrees well with the critical time identified earlier from our numerical simulations.  We have included this characteristic as the solid line in Figure~\ref{fig4}.

\section{Late stage of the collapse: numerical results}
\label{late_times}

\subsection{Numerical Methods}

We performed numerical simulations of OS collapse in moving-puncture coordinates with two separate and independent codes, producing consistent results.

One code was the 3D code developed by the Illinois group (see, e.g.~\cite{farris11,paschalidis11,etienne10,etienne09}).   The code adopts the BSSN formalism \cite{ShiN95,BauS99} to evolve the gravitational fields and a high-resolution shock-capturing scheme to evolve the relativistic hydrodynamics.  In order to achieve sufficient numerical resolution the code adopts adaptive mesh refinement.  For the results shown in this paper we used 8 levels of refinement, resulting in a resolution of $\delta X / M = 0.0078125$ on the finest level with the outer boundaries imposed at $\pm 64M$.

Our other code is a 1D code, originally developed for vacuum spacetimes in Ref.~\cite{Bro09}, that uses the Cartoon method to impose spherical symmetry.   We have supplemented this code with a simple finite-difference implementation of the equations of relativistic hydrodynamics assuming dust.  All of the results shown in this Section were obtained with the 3D code, and have been confirmed with the 1D code. 

We also show results for two different shift conditions.  The first condition is the gamma-driver condition commonly used in moving-puncture simulations 
\begin{equation} \label{gamma_driver}
\begin{array}{rcl}
\partial_t \beta^i & = & \displaystyle \beta^k \partial_k \beta^i + \frac{3}{4} B^i \\[3mm]
\partial_t B^i & = & \displaystyle \beta^k \partial_k \beta^i + (\partial_t \bar \Gamma^i)_{\rm rhs} - 
	\beta^k \partial_k \bar \Gamma^i - \eta B^i
\end{array}
\end{equation}
Here the auxiliary quantity $B^i$ has been introduced in order to turn an otherwise second-order in time equation into a first-order equation, the conformal connection functions $\bar \Gamma^i \equiv \bar \gamma^{lm} \bar \Gamma^i_{lm}$ are defined as contractions of the Christoffel symbols associated with the conformally related spatial metric $\bar \gamma_{ij}$, and $(\partial_t \bar \Gamma^i)_{\rm rhs}$ denotes their time derivative (see, e.g., Box 11.1 in \cite{BauS10}).   Finally, we will adopt $\eta = 2/M$ for all results shown here.

We also consider the alternative condition
\begin{equation} \label{alt_shift}
\partial_t \beta^i = \mu_S \bar \Gamma^i - \eta \beta^i + \beta^j \partial_j \beta^i.
\end{equation}
which is equivalent to the traditional condition (\ref{gamma_driver}) when $\mu_S = 3/4$ \cite{MetBKC06}.  Thierfelder {\em et al} \cite{ThiBHBR11} considered two cases in their analysis of fluid stars: $\mu_S = 1$,  and $\mu_S = \alpha^2$. We also consider the condition (\ref{alt_shift}) below; whenever we do, we adopt $\mu_S = \alpha^2$. 

\begin{figure}
\includegraphics[width=3in]{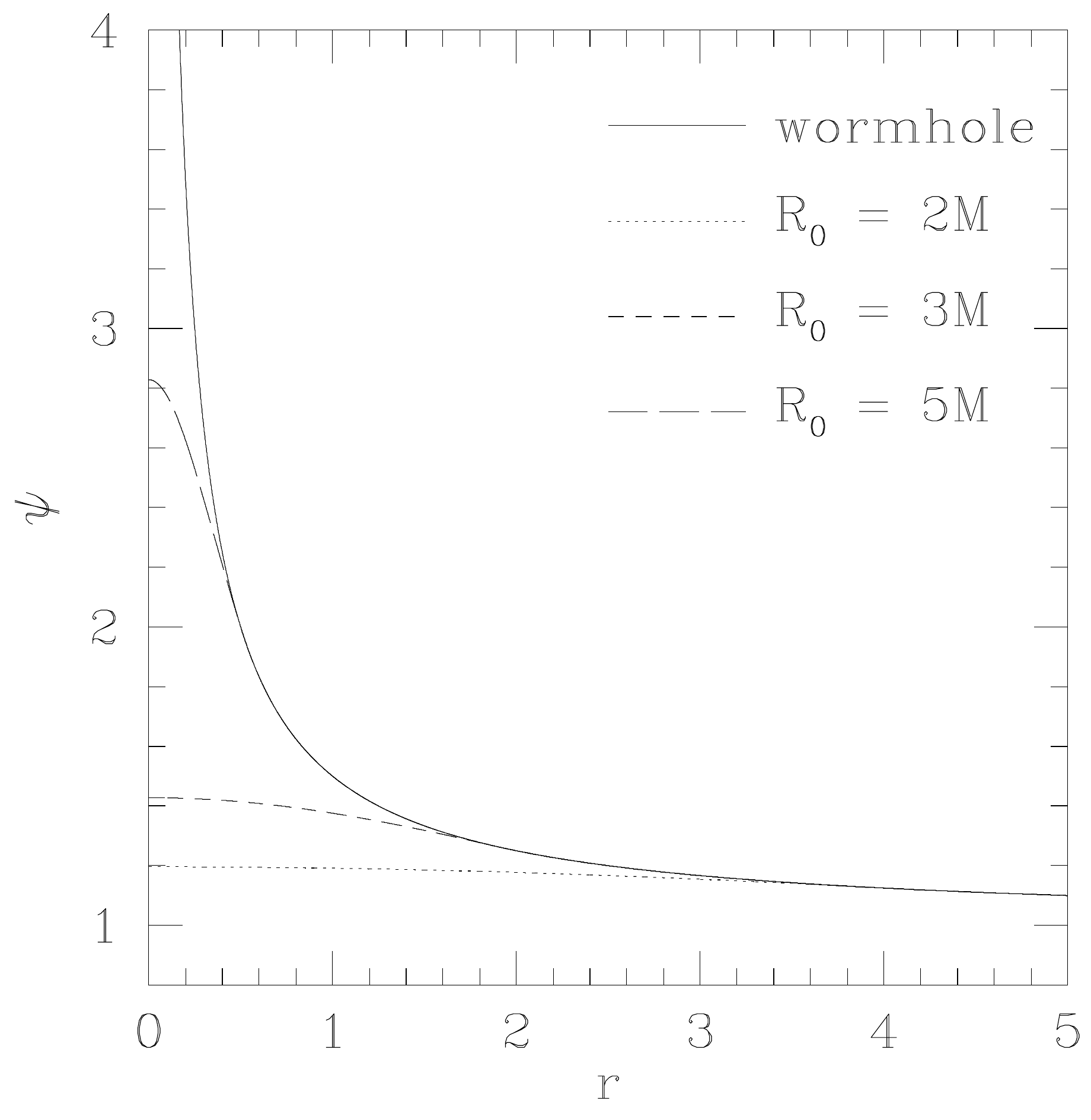}
\caption{Conformal factor $\psi$ as a function of isotropic radius $r$ for the initial dust sphere. The curves correspond to $R_0 = 5M$, $R_0 = 3M$ and $R_0 = 2M$; also included is the conformal factor $\psi = 1 + M/(2r)$ for ``wormhole" Schwarzschild data, which diverges for $r \rightarrow 0$.   The dust curves match the wormhole curve at $r_0 = 3.94M$ (for $R_0 = 5M$), $r_0 = 1.87M$ (for $R_0 = 3M$) and $r_0 = 0.5M$ (for $R_0 = 2M$).}
\label{fig4a}
\end{figure}

Initial data for our codes, which adopt cartesian coordinates, can be constructed by bringing the metric into isotropic coordinates.  The exterior Schwarzschild metric then takes the familiar form
\begin{equation}
dl^2 = \left(1 + \frac{M}{2r} \right)^4 (dr^2 + r^2 d\Omega^2),
\end{equation}
where $dl^2$ denotes a spatial line element, and where $r$ is an isotropic radius.  In the interior, we can transform the Friedmann metric (\ref{met_gn}) to isotropic coordinates by letting $\chi = 2\arctan(c r)$ with the constant $c$ chosen such that the conformal factors in the interior and exterior match at the star's surface.   We then obtain the initial (spatial) line element 
\begin{equation}
dl^2 = \psi^4(dr^2 + r^2 d\Omega^2)
\end{equation} 
with 
\begin{equation}
    \psi = \left\{ \begin{array}{ll} 
    \displaystyle \left( \frac{\left(1 + \sqrt{1 - 2M/R_0} \right)r_0 R_0^2}
    {2 r_0^3 + Mr^2}\right)^{1/2} \ ,  & r \le r_0 \ , \\[5mm]
    \displaystyle 1 + \frac{M}{2r}  \ , & r > r_0 \ , 
    \end{array} \right. 
\end{equation}
where $r_0 = R_0 \left(1 - M/R_0 + \sqrt{1 - 2M/R_0}\right)/2$. The initial data also include $K_{ij} = 0$.   Profiles of the conformal factor $\psi$ for different values of $R_0$ are shown in Figure~\ref{fig4a}.

\subsection{Results for $R_0 = 5M$}

\begin{figure}
\includegraphics[width=5in]{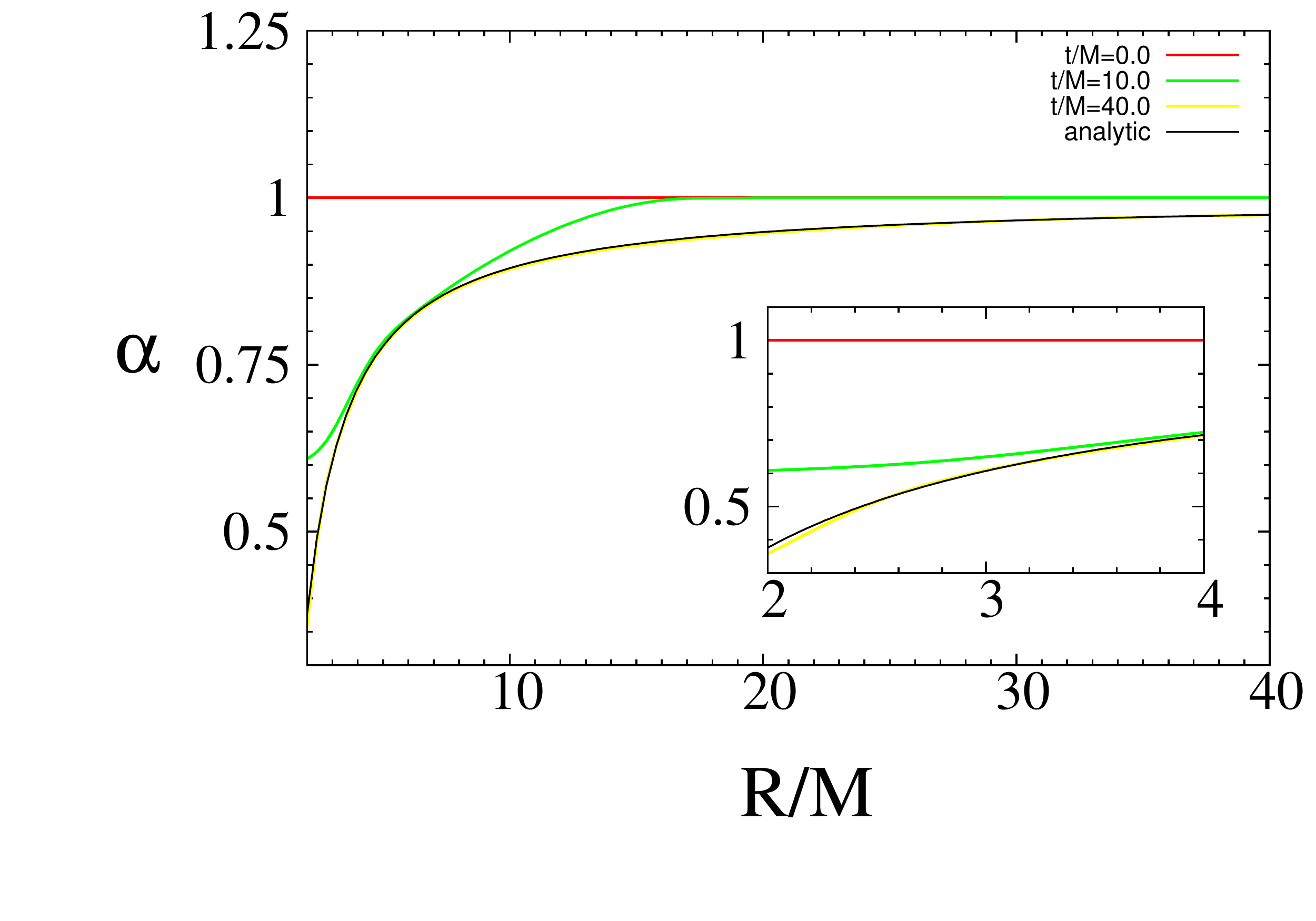}
\caption{Profiles of the lapse $\alpha$ as a function of radius at different times $t$, as obtained with the shift condition (\ref{gamma_driver}). Red, green and yellow lines correspond to $t/M=$ 0, 10, and 40, respectively. At late times, the lapse asymptotes to that the lapse of a  trumpet slice of Schwarzschild in ``stationary 1+log" slicing", which is included as a black line (see \cite{HanHOBO08}).}
\label{fig5}
\end{figure}

 \begin{figure}
\includegraphics[width=6in]{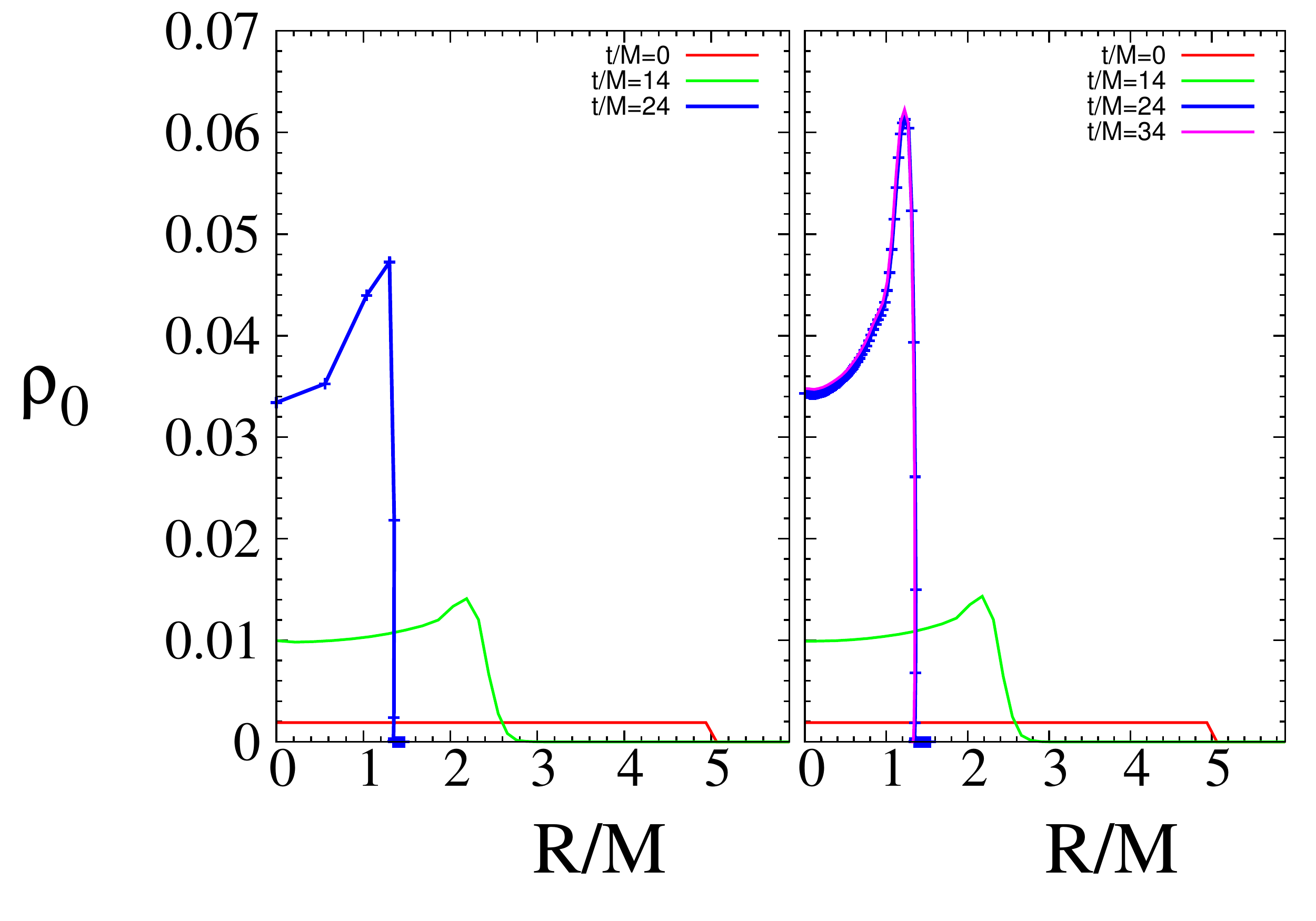}
\caption{Profiles of the density $\rho_0$ as a function of areal radius $R$ at different times $t$.  The left panel shows result for the shift condition (\ref{gamma_driver}), while the right panel shows results for the shift condition (\ref{alt_shift}).  For the former condition (left panel), grid-stretching leads to all grid points being pushed out of the star at late times, leaving the star under-resolved, and making the evolution of the density very unreliable.  To highlight this effect we marked individual gridpoints with crosses for the inner regions of our spacetime at $t/M$ = 24.  For the latter condition (right panel), grid points are frozen inside the star, so that it remains resolved even at late times. Red, green, blue and magenta lines correspond to $t/M=$ 0, 14, 24 and 34 respectively. Note that there is no line corresponding to $t.M=34$ in the left panel, as the matter is completely unresolved at this time.}
\label{fig6}
\end{figure}

\begin{figure}
\includegraphics[width=6in]{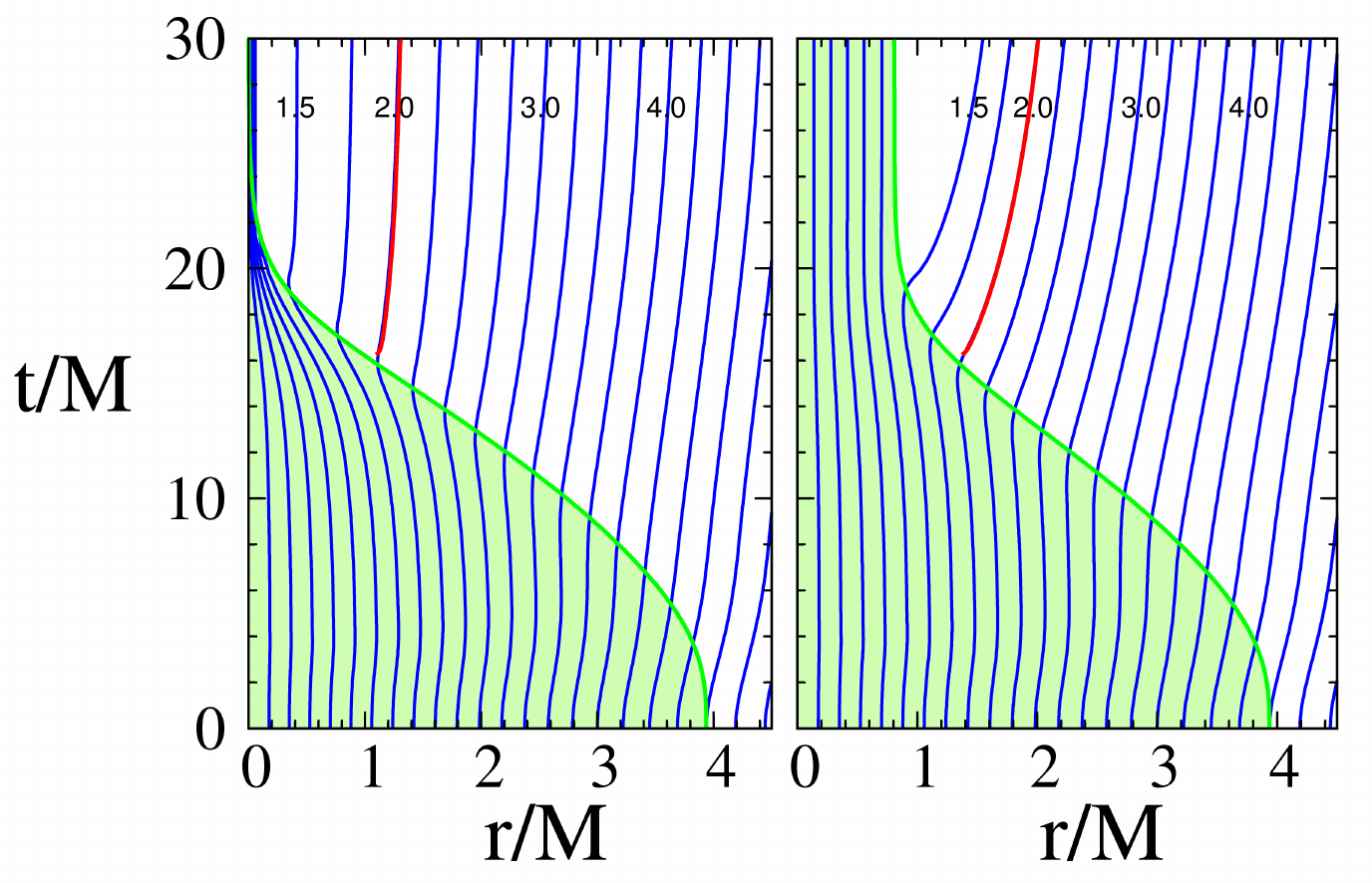}
\caption{Spacetime diagrams for OS collapse (compare Fig.~6 in \cite{ThiBHBR11}).  The left panel shows results for the shift condition (\ref{gamma_driver}), while the right panel shows results for the shift condition (\ref{alt_shift}).  Blue lines correspond to lines of constant areal radius.  The green region marks the stellar interior.  The red line denotes the apparent horizon which, as in Gaussian normal coordinates, first appears on the stellar surface.}
\label{STdiagram}
\end{figure}

\begin{figure}
\includegraphics[width=6in]{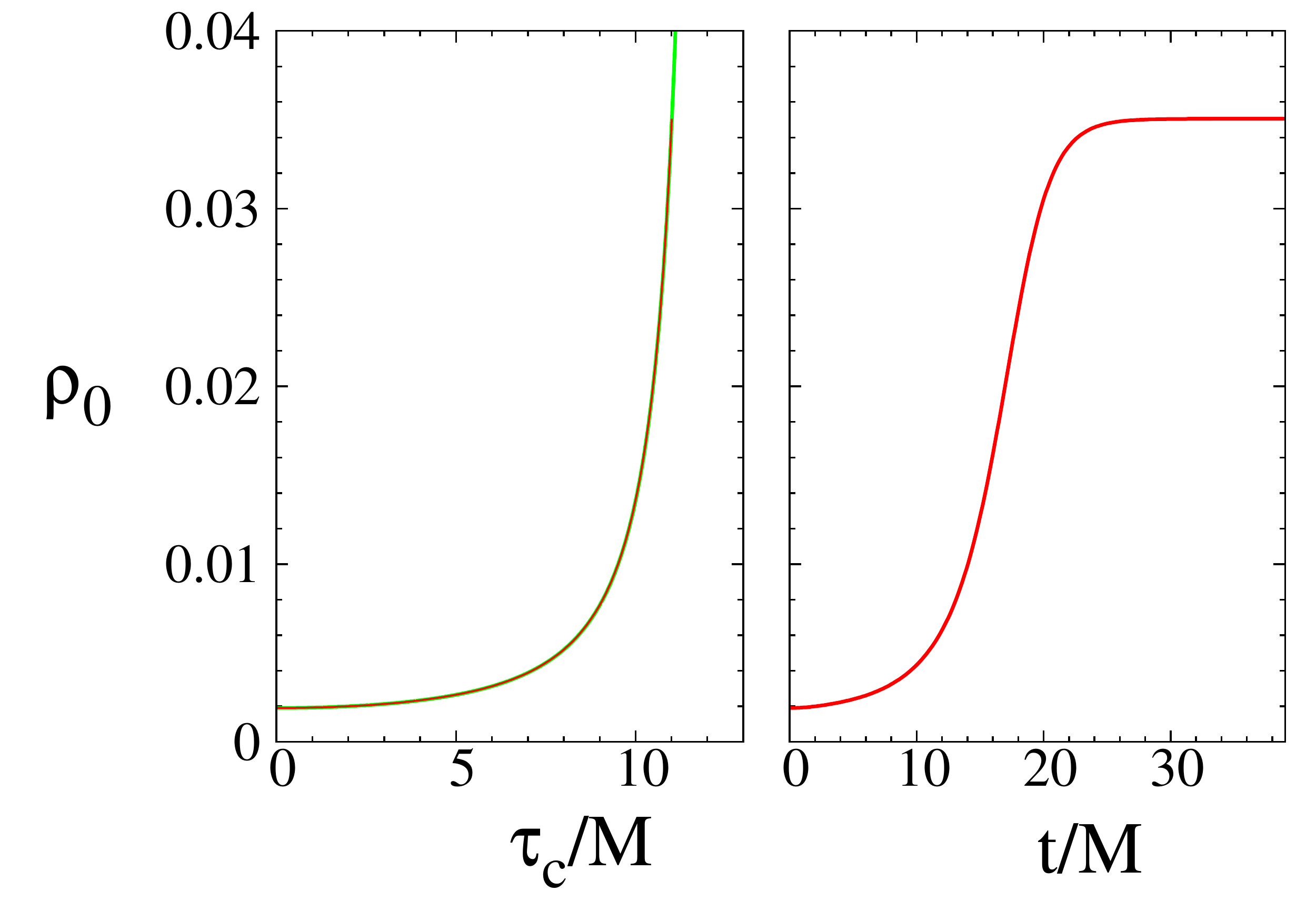}
\caption{The central density $\rho_0^c$ as a function of time, obtained with the shift condition (\ref{alt_shift}) with $\mu_S = \alpha^2$.  The left panel shows the central density as a function of proper time $\tau$.  Also included is the analytic result (\ref{density}).  The numerical solution terminates at the proper time $\tau_{\rm limit}$, when the lapse collapses to zero, preventing any further advance in proper time $d\tau = \alpha dt$.  The right panel shows the central density as a function of coordinate time $t$.  In this graph the density levels off to the constant value $\rho_0^{\rm limit}$, indicating that, while coordinate time continues to advance, proper time does not. In each plot, the numerical value is shown in red, while in the left panel the analytic value is also shown in green.}
\label{fig7}
\end{figure}

\begin{figure}
\includegraphics[width=6in]{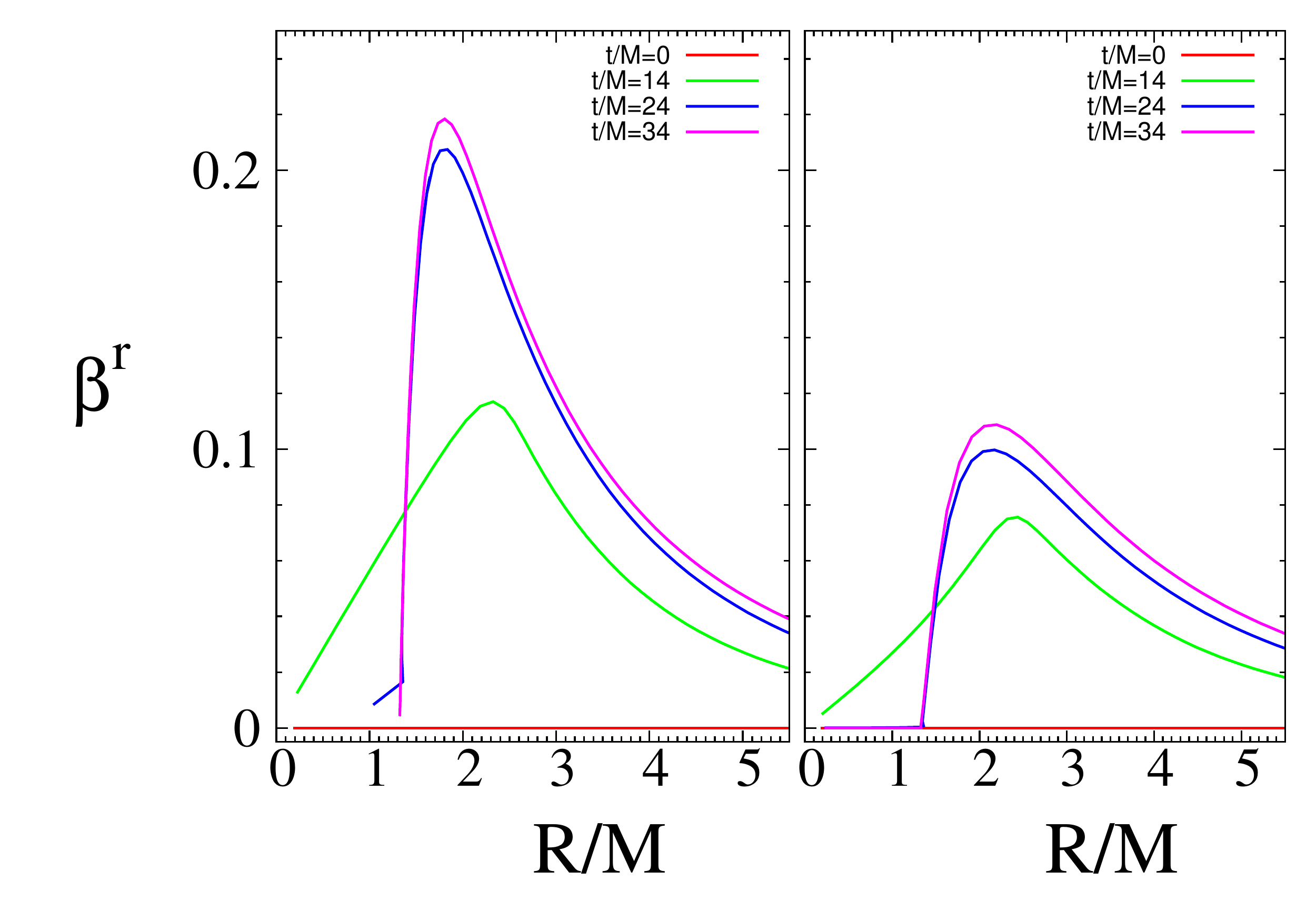}
\caption{Profiles of the shift at different coordinate times $t$.  The left panel shows values of the shift as obtained with the shift condition (\ref{gamma_driver}), while the right panel shows results for the condition (\ref{alt_shift}). Red, green, blue and magenta lines correspond to $t/M=$ 0, 14, 24 and 34, respectively.}
\label{fig8}
\end{figure}

Results for the lapse $\alpha$ at early times have already been shown in Figure~\ref{fig2}.  All of these results have been obtained with the gauge condition (\ref{gamma_driver}), and demonstrate the existence of a region of decreasing radius around the origin in which these quantities remain spatially constant.  We now turn to results for late times of the collapse.  In this Section we will focus on $R_0 = 5M$; we will discuss more general results below.

In Figure \ref{fig5} we show profiles of the lapse as a function of radius at different times $t$, as obtained with the shift condition (\ref{gamma_driver}).  We also include the analytical lapse for trumpet slices of Schwarzschild in ``stationary 1+log" slicing, which satisfies the equation
\begin{equation}
\alpha^2 = 1 - \frac{2M}{R} + \frac{C^2 e^\alpha}{R^4}
\end{equation}
with $C = (\sqrt{2}/16) (\sqrt{10} + 3)^{3/2} e^{(3 - \sqrt{10})/2} M^2 \approx 1.24672 M^2$ (see, e.g., \cite{HanHOBO08}).  Evidently, the lapse as obtained from OS collapse in moving-puncture coordinates asymptotes to that for a Schwarzschild black hole.  This is not surprising for the stellar exterior, of course, but one might have expected that an interior region remains in which the density remains different from zero, and in which the lapse differs from that of a vacuum spacetime.  As discussed in \cite{ThiBHBR11}, however, the shift condition (\ref{gamma_driver}) pushes all grid points into the stellar exterior in a finite time, so that the stellar interior remains unresolved and removed from the grid.

This effect is also visible in Figure \ref{fig6}, where we show profiles of the density $\rho_0$ for different coordinate times $t$.    In the left panel we again show results for the shift condition (\ref{gamma_driver}).  At late times, the density drops to zero everywhere, indicating that  all grid points having been pushed out of the stellar interior.  In the right panel we show the same results for the shift condition (\ref{alt_shift}).  Here, grid points are frozen at late times, and remain in the stellar interior, which is well-resolved even at late times.

In Figure \ref{STdiagram} we show spacetime diagrams, similar to those in Fig.~6 of \cite{ThiBHBR11}, to further illustrate the influence of the shift condition on the late-time trajectories of gridpoints.  In each plot, the blue lines correspond to surfaces of constant areal radius, the green region marks the stellar interior, and the red line denotes the apparent horizon.  In the left panel we show results for the shift condition (\ref{gamma_driver}), for which no gridpoints remain in the stellar interior at late times, while the right panel shows results for the shift condition (\ref{alt_shift}), for which the stellar interior remains well resolved even at late times.  The apparent horizon first appears when the stellar surface passes through $R = 2M$, as in Gaussian-normal coordinates (compare, e.g., Fig.~1.3 in \cite{BauS10}).

In Figure \ref{fig7} we show the central density $\rho_0^c$ as a function of time, as obtained with the shift condition (\ref{alt_shift}).  In the left panel we show the central density as a function of proper time, and compare with the analytical result (\ref{density}).   We see that the numerical evolution terminates at the proper time $\tau_{\rm limit}$ (see also the spacetime diagram in Figure \ref{fig4}), when the lapse collapses to zero and prevents any further advance in proper time (recall $d \tau = \alpha dt$).  In the right panel we show the central density as a function of coordinate time, which continues to advance even after the lapse has collapsed to zero.  In this graph, the density levels off to the limiting value $\rho_0^{\rm limit}$, corresponding to the value of the central density at proper time $\tau_{\rm limit}$.

Finally, we show profiles of the shift $\beta^i$ in Figure \ref{fig8}.   As before, we show results obtained with the shift condition (\ref{gamma_driver}) in the left panel.  Note that at late times, these lines terminate at a finite value of the areal radius $R$, corresponding to the limiting surface of the trumpet solution.  In the right panel, showing results obtained with (\ref{alt_shift}), the shift for small $R$ drops to zero at late times, indicating that the corresponding gridpoints are ``frozen" in the interior of the star.


\subsection{Effect of $R_0$}

In the previous Section we showed that for OS collapse with an initial areal radius of $R_0 = 5M$, the 1+log slicing condition successfully halts the advance of the spatial slices before the singularity forms.  In this Section we consider other values of $R_0$; in particular we evaluate whether or not, for smaller initial radii, the singularity forms too quickly for the lapse to halt the evolution.  
 
We have investigated this question numerically.   Our simulations show that the 1+log condition has no difficulty handling the collapse of arbitrarily small dust spheres, which, perhaps, is not surprising.  Figure \ref{fig4a} shows the initial conformal factor $\psi$ as a function of isotropic radius $r$ for different initial areal radii $R_0/M$, as well as the conformal factor for ``wormhole'' Schwarzschild initial data.  As $R_0$ decreases and approaches the limit $2M$, the initial data approach that of a black hole.  The black hole is the most stringent scenario, for which the lapse must collapse as quickly as possible to keep the slice from reaching the singularity.  We also see this from the proper time (\ref{tau_sing}) that elapses before the singularity is reached.  For $R_0 = 2M$ we have $\tau_{\rm sing} = \pi M$, which is the same time required for a normal observer at the center of a Kruskal diagram of a Schwarzschild black hole to reach the singularity.   Numerous numerical simulations of the Schwarzschild spacetime, evolving ``wormhole" initial data with moving-puncture coordinates, demonstrate that the lapse collapses sufficiently fast to avoid the singularity (e.g.~\cite{HanHPBO06}); apparently the same holds for OS collapse for any value of $R_0$.  

For $R_0 = 2M$, the initial ball of dust is just inside its own Schwarzschild radius. In the spirit of black hole ``stuffing" \cite{EtiFLSB07,BroSSTDHP07,Brown:2008sb}, we can evolve the data with or without the matter fields.  Omitting the matter terms in the black hole interior and evolving the initial data as if they were vacuum data leads to a violation of the constraints in the black hole interior; however, these constraint violations do not affect the black hole exterior, and disappear even in the interior after a short evolution of the data.
  
\section{Summary}
\label{sum}

We analyze OS collapse, describing the collapse of a homogeneous dust sphere initially at rest to a black hole, in moving-puncture coordinates. These coordinates have proven particularly robust for handling vacuum black holes in numerical simulations and this has motivated our interest in exploring how they behave in the presence of matter sources.   

We obtain the complete solution for OS collapse numerically and elucidate some of the key results analytically, especially for the early-time behavior. In particular, we show that if the initial lapse $\alpha_0$ is chosen constant, then the lapse $\alpha$, the density $\rho$ and the mean curvature $K$ will remain spatially constant in a region around the center of the star that decreases in size, and that disappears at a critical time $\tau_{\rm gauge}$.  In this region, which is marked as the shaded region in Figure \ref{fig4},  the lapse is given by equation (\ref{lapse_const}).   The region is limited by a gauge mode that originates at the surface of the star and travels inwards at a speed that exceeds the speed of light (see equation (\ref{v_prop})).  We identify $\tau_{\rm gauge}$ with the time at which this characteristic, marked by the solid line in Figure \ref{fig4}, reaches the stellar center.  At the center of the star, equation (\ref{lapse_ll}) serves as a lower limit on the lapse; the lapse agrees with this lower limit for times up to $\tau_{\rm gauge}$, but starts to deviate abruptly at that time.  

With the gamma-driver shift condition (\ref{gamma_driver}), grid points are pulled out of the stellar interior at late times, so that the numerical grid covers only the exterior of the star. The numerical solution asymptotes to a trumpet slice of a Schwarzschild spacetime.  This confirms the findings of \cite{ThiBHBR11} and demonstrates that, even in the absence of pressure, the lapse collapses sufficiently fast to avoid the forming singularity, and to allow the transition to a trumpet geometry.  We also confirm that with the alternative shift condition (\ref{alt_shift}) with $\mu_S = \alpha^2$ the coordinate grid points remain in the stellar interior and the star is well-resolved even at late times.  

\ackn

ANS gratefully acknowledges support through a Clare Boothe Luce Fellowship for undergraduate students.  BF gratefully acknowledges support through NASA Earth and Space Science Fellowship NNX09AO64H.  This work was supported in part by NSF grant PHY-0756514 to Bowdoin College, NSF grant PHY-0758116 to North Carolina State University, and NSF grant PHY-0963136  and NASA grant NNX10AI73G to the University of Illinois at Urbana-Champaign.


\section*{References}
 \begin{thebibliography}{10}

\bibitem{Pre05b}
F.~{Pretorius}.
\newblock Evolution of binary black-hole spacetimes.
\newblock {\em \prl} {\bf 95}, 121101 (2005).

\bibitem{BakCCKM06a}
J.~G. {Baker}, J.~{Centrella}, D.-I. {Choi}, M.~{Koppitz}, and J.~{van Meter}.
\newblock Gravitational-wave extraction from an inspiraling configuration of
  merging black holes.
\newblock {\em \prl}  {\bf 96}, 111102 (2006).

\bibitem{CamLMZ06}
M.~{Campanelli}, C.~O. {Lousto}, P.~{Marronetti}, and Y.~{Zlochower}.
\newblock Accurate evolutions of orbiting black-hole binaries without excision.
\newblock {\em \prl} {\bf 96}, 111101 (2006).

\bibitem{ShiN95}
M.~{Shibata} and T.~{Nakamura}.
\newblock {Evolution of three-dimensional gravitational waves: Harmonic slicing
  case}.
\newblock {\em \prd} {\bf 52}, 5428 (1995).

\bibitem{BauS99}
T.~W. {Baumgarte} and S.~L. {Shapiro}.
\newblock {Numerical integration of Einstein's field equations}.
\newblock {\em \prd} {\bf 59}, 024007 (1999).

\bibitem{BauS10}
T.~W. {Baumgarte} and S.~L. {Shapiro}.
\newblock {\em Numerical Relativity: Solving Einstein's Equations on the
  Computer}.
\newblock Cambridge University Press, Cambridge, 2010.

\bibitem{BonMSS95}
C.~{Bona}, J.~{Mass{\'o}}, E.~{Seidel}, and J.~{Stela}.
\newblock {New Formalism for Numerical Relativity}.
\newblock {\em \prl} {\bf 75}, 600 (1995).

\bibitem{AlcBDKPST03}
M.~{Alcubierre}, B.~{Br{\"u}gmann}, P.~{Diener}, M.~{Koppitz}, D.~{Pollney},
  E.~{Seidel}, and R.~{Takahashi}.
\newblock {Gauge conditions for long-term numerical black hole evolutions
  without excision}.
\newblock {\em \prd} {\bf 67}, 084023 (2003).

\bibitem{HanHPBO06}
M.~{Hannam}, S.~{Husa}, D.~{Pollney}, B.~{Bruegmann}, and N.~{O'Murchadha}.
\newblock {Geometry and Regularity of Moving Punctures}.
\newblock {\em \prl} {\bf 99}, 241102 (2007).

\bibitem{HanHOBGS06}
M.~{Hannam}, S.~{Husa}, N.~{\'O} {Murchadha}, B.~{Br{\"u}gmann}, J.~A.
  {Gonz{\'a}lez}, and U.~{Sperhake}.
\newblock {Where do moving punctures go?}
\newblock {\em {J.~Phys.~Conf.~Series}}  {\bf 66}, 012047 (2007).

\bibitem{Bro08}
J.~D. Brown.
\newblock {Puncture Evolution of Schwarzschild Black Holes}.
\newblock {\em \prd} {\bf 77}, 044018 (2008).

\bibitem{HanHOBO08}
M.~{Hannam}, S.~{Husa}, F.~{Ohme}, B.~{Br{\"u}gmann}, and N.~{\'O} {Murchadha}.
\newblock Wormholes and trumpets: the {S}chwarzschild spacetime for the
  moving-puncture generation.
\newblock {\em \prd} {\bf 78}, 064020 (2008).

\bibitem{DenWBB10}
K.~A. {Dennison}, J.~P. {Wendell}, T.~W. {Baumgarte}, and J.~D. {Brown}.
\newblock Trumpet slices of the {S}chwarzschild-{T}angherlini spacetime.
\newblock {\em \prd} {\bf 82}, 124057 (2010).

\bibitem{FabBEST07}
J.~A. {Faber}, T.~W. {Baumgarte}, Z.~B. {Etienne}, S.~L. {Shapiro}, and
  K.~{Taniguchi}.
\newblock {Relativistic hydrodynamics in the presence of puncture black holes}.
\newblock {\em \prd} {\bf 76}, 104021 (2007).

\bibitem{BaiR06}
L.~{Baiotti} and L.~{Rezzolla}.
\newblock Challenging the paradigm of singularity excision in gravitational
  collapse.
\newblock {\em \prl} {\bf 97}, 121101 (2006).

\bibitem{MonFS08}
P.~J. {Montero}, J.~A. {Font}, and M.~{Shibata}.
\newblock {Nada: A new code for studying self-gravitating tori around black
  holes}.
\newblock {\em \prd} {\bf 78}, 064037 (2008).

\bibitem{etienne09}
Z.~B. {Etienne}, Y.~T. {Liu}, S.~L. {Shapiro}, and T.~W. {Baumgarte}.
\newblock {General relativistic simulations of black-hole-neutron-star mergers:
  Effects of black-hole spin}.
\newblock {\em \prd} {\bf 79}, 044024 (2009).

\bibitem{ThiBB11}
M.~{Thierfelder}, S.~{Bernuzzi}, and B.~{Br\"ugmann}.
\newblock Numerical relativity simulations of binary neutron stars.
\newblock {\em \prd} {\bf 84}, 044012 (2011).

\bibitem{ThiBHBR11}
M.~{Thierfelder}, S.~{Bernuzzi}, D.~{Hilditch}, B.~{Br\"ugmann}, and
  L.~{Rezzolla}.
\newblock The trumpet solution from spherical gravitational collapse with
  puncture gauges.
\newblock {\em \prd} {\bf 83}, 064022 (2011).

\bibitem{MetBKC06}
J.~R. {van Meter}, J.~G. {Baker}, M.~{Koppitz}, and D.-I. {Choi}.
\newblock How to move a black hole without excision: gauge conditions for the
  numerical evolution of a moving puncture.
\newblock {\em \prd} {\bf 73}, 124011 (2006).

\bibitem{Shi99a}
M.~{Shibata}.
\newblock Fully general relativistic simulation of merging binary clusters:
  Spatial gauge condition.
\newblock {\em Prog. Theor. Phys.} {\bf 101}, 1199 (1999).

\bibitem{OppS39}
J.~R. {Oppenheimer} and H.~{Snyder}.
\newblock {On Continued Gravitational Contraction}.
\newblock {\em \pr} {\bf 56}, 455 (1939).

\bibitem{PetST85}
L.~I. {Petrich}, S.~L. {Shapiro}, and S.~A. {Teukolsky}.
\newblock {Oppenheimer-Snyder collapse with maximal time slicing and isotropic
  coordinates}.
\newblock {\em \prd} {\bf 31}, 2459 (1985).

\bibitem{PetST86}
L.~I. {Petrich}, S.~L. {Shapiro}, and S.~A. {Teukolsky}.
\newblock {Oppenheimer-Snyder collapse in polar time slicing}.
\newblock {\em \prd}  {\bf 33}, 2100 (1986).

\bibitem{BauST95}
T.~W. {Baumgarte}, S.~L. {Shapiro}, and S.~A. {Teukolsky}.
\newblock {Computing supernova collapse to neutron stars and black holes}.
\newblock {\em \apj} {\bf 443}, 717 (1995).

\bibitem{Alc03}
M.~{Alcubierre}.
\newblock Hyperbolic slicings of spacetime: singularity avoidance and gauge
  shocks.
\newblock {\em \cqg} {\bf 20}, 607 (2003).

\bibitem{Beyer:2004sv}
H.~R. Beyer and O.~Sarbach.
\newblock {On the well posedness of the Baumgarte-Shapiro-Shibata-Nakamura
  formulation of Einstein's field equations}.
\newblock {\em \prd} {\bf 70}, 104004 (2004).

\bibitem{Bro09}
J.~D. Brown.
\newblock Probing the puncture for black hole simulations.
\newblock {\em \prd} {\bf 80}, 084042 (2009).

\bibitem{farris11}
B.~D. {Farris}, Y.~T. {Liu}, and S.~L. {Shapiro}.
\newblock {Binary black hole mergers in gaseous disks: Simulations in general
  relativity}.
\newblock {\em ArXiv}:1105.2821 (2011).

\bibitem{paschalidis11}
V.~{Paschalidis}, Z.~{Etienne}, Y.~T. {Liu}, and S.~L. {Shapiro}.
\newblock {Head-on collisions of binary white dwarf-neutron stars: Simulations
  in full general relativity}.
\newblock {\em \prd} {\bf 83}, 064002 (2011).

\bibitem{etienne10}
Z.~B. {Etienne}, Y.~T. {Liu}, and S.~L. {Shapiro}.
\newblock {Relativistic magnetohydrodynamics in dynamical spacetimes: A new
  adaptive mesh refinement implementation}.
\newblock {\em \prd} {\bf 82}, 084031 (2010).

\bibitem{EtiFLSB07}
Z.~B. {Etienne}, J.~A. {Faber}, Y.~T. {Liu}, S.~L. {Shapiro}, and T.~W.
  {Baumgarte}.
\newblock {Filling the holes: Evolving excised binary black hole initial data
  with puncture techniques}.
\newblock {\em \prd} {\bf 76}, 101503 (2007).

\bibitem{BroSSTDHP07}
J.~D. {Brown}, O.~{Sarbach}, E.~{Schnetter}, M.~{Tiglio}, P.~{Diener},
  I.~{Hawke}, and D.~{Pollney}.
\newblock {Excision without excision}.
\newblock {\em \prd} {\bf 76}, 081503 (2007).

\bibitem{Brown:2008sb}
J.~D. Brown, P.~Diener, O.~Sarbach, E.~Schnetter, and M.~Tiglio.
\newblock {Turduckening black holes: An Analytical and computational study}.
\newblock {\em \prd} {\bf 79}, 044023 (2009).

\end{thebibliography}
 \end{document}